\documentclass[a4paper]{article}
\usepackage{cite, graphicx, subfigure, amsmath, appendix} 
\usepackage{setspace}
\usepackage{fullpage}
\usepackage{fancyhdr}
\usepackage{fullpage}
\usepackage{caption} 
\usepackage{hyperref}
\usepackage{multirow}
\usepackage{comment}
\usepackage{wrapfig}
\usepackage{amssymb}
\usepackage{amsfonts}
\usepackage{hyperref}

\date{}

\begin{document}
\newcommand\bra[2][]{#1\langle {#2} #1\rvert}
\newcommand\ket[2][]{#1\lvert {#2} #1\rangle}
\title{Loading Probability Distributions in a Quantum circuit}
\author{Kalyan Dasgupta$^1$ \and Binoy Paine$^2$}
\date{%
    $^1$IBM Research, Bangalore, India\\%
    $^2$Indian Institute of Science, Bangalore, India\\[2ex]%
    }
\maketitle
\begin{abstract}
Quantum circuits generating probability distributions has applications in several areas. Areas like finance require quantum circuits that can generate distributions that mimic some given data pattern. Hamiltonian simulations require circuits that can initialize the wave function of a physical quantum system. These wave functions, in several cases, are identical to some very well known probability distributions. In this paper we discuss ways to construct parameterized quantum circuits that can generate both symmetric as well as asymmetric distributions. We follow the trajectory of quantum states as single and two qubit operations get applied to the system, and find out the best possible way to arrive at the desired distribution. The parameters are optimized by a variational solver. We present results from both simulators as well as real IBM quantum hardwares.
\end{abstract}
\section{Introduction}
Generating probability distributions in a quantum circuit could have many uses in different application areas. One area where it has seen use cases is in the area of finance \cite{Zoufal}. Generating probability distributions, essentially, entails assigning probability amplitudes to basis states of a quantum system. A system with $n$ qubits would have $2^n$ basis states. The generator of the probability distribution has to assign amplitudes to these basis states in such a manner that the results of the measurement statistics closely resemble a given probability distribution. Another use case of such generators is in creating initial states in Hamiltonian simulation. A Hamiltonian simulation requires an initial wave function over which the time evolution could be simulated \cite{SebastianYT}, \cite{Mike_Ike}. In this case, however, it is more of a probability amplitude generator, that is needed. We can make suitable changes to our probability distribution generators to get a desired probability amplitude distribution. 

\subsection{Related Literature}
\begin{figure}[h]
\centering 
\includegraphics[width=4.5in]{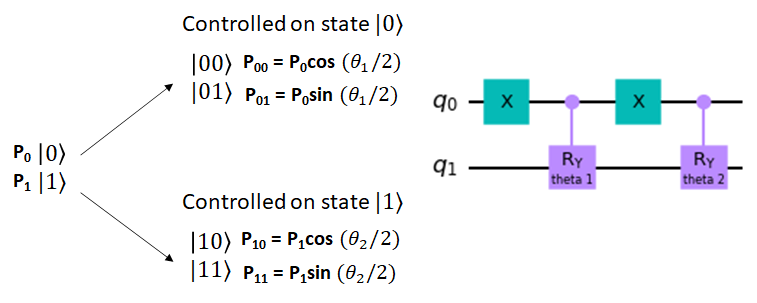} 
\caption{Probability distribution in a 2 qubit system}
\label{fig:grover_exp}
\end{figure}

One can find some papers written on generating probability distributions. One of the earliest papers on generating probability distributions is by \cite{Grover}. In this paper, the author gives a scheme on how to generate a superposition of quantum states of the form given in (\ref{grover_sup1}), that resembles a given probability distribution.
\begin{flalign}
\ket{\psi} = \sum_i \sqrt{p_i}\ket{i} \label{grover_sup1}
\end{flalign}
The method followed in this scheme is an incremental expansion scheme, whereby, at every point in the expansion, the system is expanded by one qubit by a controlled rotation. The existing qubits form the control set and the target qubit is the newly added qubit. This can be better understood by considering that every state of the existing system will expand by the addition (tensor product) of a $\ket{0}$ and a $\ket{1}$ to its left as shown in equation (\ref{grover_sup2}).
\begin{flalign}
\sqrt{p_i}\ket{i} \rightarrow \sqrt{p_i}\ket{i} \otimes \left[ \cos{\theta}\ket{0} + \sin{\theta}\ket{1} \right] \label{grover_sup2}.
\end{flalign}
Fig. \ref{fig:grover_exp} illustrates the expansion scheme by taking the example of a transition from a 1 qubit system to a 2 qubit system. This method requires $2^{n-1}$ controlled $R_Y$ gates and $\approx2^n$ $X$ gates, for a $n$ qubit system.
In Fig. \ref{fig:grover_exp}, the variables $P_0, P_1, P_{00}$, etc., denote the probability amplitudes and not the probabilities. A more detailed explanation can be found in \cite{Nakamura}.

\begin{figure}[h]
\centering 
\includegraphics[width=3.5in]{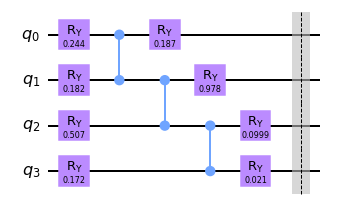} 
\caption{Probability distribution generation using variational circuits}
\label{fig:Zoufal_var}
\end{figure}

In \cite{Zoufal}, the authors use a Quantum Generative Adversarial Networks (QGANs) to learn distributions from given training data. The algorithm uses a hybrid quantum-classical computation approach, where a quantum computer generates a probability distribution and a classical computer tries to discriminate between the generated distribution and the distribution of the training data. The quantum computer generates a probability distribution using variational circuits. The variational circuit uses $R_Y$ rotation gates and controlled $Z$ ($CZ$) gates for entanglement. Fig. \ref{fig:Zoufal_var} shows one layer of the variational circuit in a 4 qubit system. 
For a $n$ qubit system, this method requires an initial set of $n$ rotation parameters and then for every layer we have $n$ set of parameters for the $n$ number of $R_Y$ gates, and $n-1$ or $n$ number of $CZ$ gates depending upon the type of entanglement (linear or circular). Overall, it requires $nl + n$ parameters for $l$ layers of the parameterized circuit. To begin with, the variational method uses a normal distribution against which the generated distribution is compared. The loss function used is $\|.\|_2$ of the difference between the generated and the desired distributions. For the discriminator the training data is sampled from log-normal distributions. 

In \cite{Plesch}, the authors give a quantum circuit scheme to prepare arbitrary quantum states. The authors were able to reduce the upper bound on the number of CNOT gates required for an even number of qubits. The authors also show that some part of circuit computation can be performed in parallel, thus reducing the computational depth. In \cite{Aharonov}, the authors discuss the evolution of quantum walks as a function of time over a graph. Under certain conditions, the limiting distribution of the evolution is uniform. 

In this article, we propose a method to generate symmetrical and asymmetrical probability distributions by studying the trajectory of the individual quantum states under the action of rotation and entanglement gates. The objective is to generate distributions that does not require circuit elements that grow exponentially with the number of qubits. The approach exploits symmetry properties of distribution curves to do away with redundant rotation and controlled gates. Variational solvers are used to fix the rotation parameters of the gates, with some constraints, such that a finer 1:1 correspondence between the probability of occurrence of a state and points in the desired probability distribution curve is obtained. The ideas given here could be used as a general framework to generate a variety of probability distribution curves.

The rest of the paper is organized as follows. In section 2, we discuss the basic concept that the methodology proposed here uses. It discusses the trajectory of a quantum state as qubits are added and unitary operations in the form of gates get applied. This section also discusses ways to generate symmetric and asymmetric (skewed) distributions. In section 3, we discuss the algorithm to generate a distribution using a variational solver. In section 4 we present some results that were obtained using both simulators and quantum hardwares. Section 5 gives the conclusions.

\section{Tracing the trajectory of quantum states}
\begin{figure}[h]
\centering 
\includegraphics[width=5in]{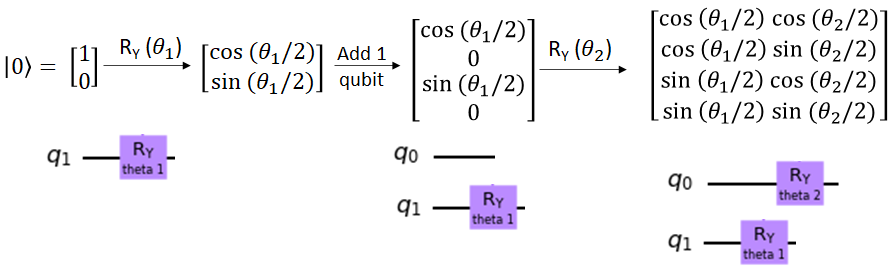} 
\caption{Effect of rotation gates on qubits}
\label{fig:hy_2qb-1}
\end{figure}

To see how rotation gates have an effect on the state of qubits, we will first take the example of a 2 qubit system. Let us say we start from a single qubit system with an initial state of $\ket{0}$. We apply a $R_Y$ gate with parameter $\theta _1$. $R_Y(\theta _1)$ gate will rotate the state from the $+Z$ axis in the bloch sphere around the $Y-Z$ plane. The expression for $R_Y(\theta _1)$ is given in (\ref{eqn:Ry}).
\begin{flalign}
R_Y (\theta _1) = \left[ \begin{array}{cc}
     \cos{(\theta _1/2)} & -\sin{(\theta _1/2)} \\
     \sin{(\theta _1/2)} &  \cos{(\theta _1/2)} \\
\end{array} \right] \label{eqn:Ry}
\end{flalign}
Thereafter, we add another qubit with an initial state of $\ket{0}$. In Qiskit, the most significant bit (MSB) comes at the bottom of the circuit \cite{qiskit}. Using that formalism, we name our first qubit as $q_1$ and the second qubit as $q_0$. Once the second qubit has been introduced, we add another rotation gate $R_Y(\theta _2)$ to the second qubit. The progression is shown in Fig. \ref{fig:hy_2qb-1}.
From Fig. \ref{fig:hy_2qb-1}, one can see a pattern emerging in the state-vector of the two qubit system. The first element of the state-vector is the probability amplitude of the binary state $\ket{00}$, the second to the state $\ket{01}$, the third to $\ket{10}$ and the fourth to $\ket{11}$. The elements containing $0$ in the MSB has a term $\cos{(\theta _1/2)}$, while the ones containing $1$, has a term $\sin{(\theta _1/2)}$. A similar thing can be said about the least significant bit with $\cos{(\theta _2/2)}$ and $\sin{(\theta _2/2)}$. We will observe a similar progression as we keep adding qubits. 

\begin{figure}[h]
\centering 
\includegraphics[width=5in]{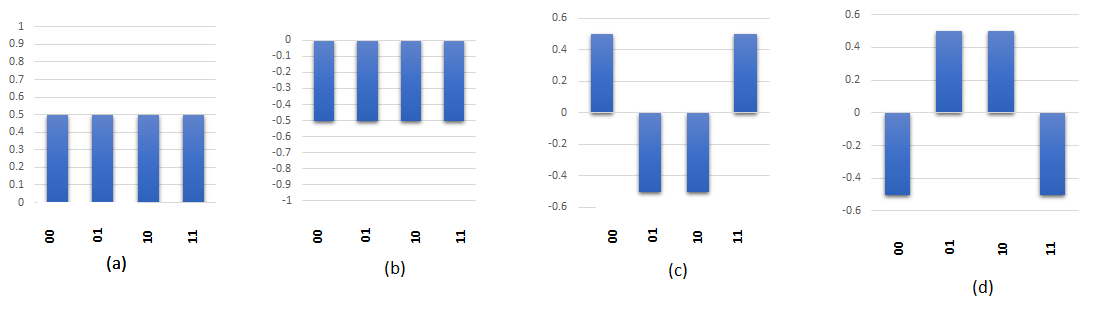} 
\caption{Symmetrical distribution of probability amplitudes under different conditions, (a) $\theta _1 = \frac{\pi}{2}, \theta _2 = \frac{\pi}{2}$, (b) $\theta _1 = \frac{\pi}{2}, \theta _2 = \frac{5\pi}{2}$, (c) $\theta _1 = \frac{3\pi}{2}, \theta _2 = \frac{3\pi}{2}$, (d) $\theta _1 = \frac{-\pi}{2}, \theta _2 = \frac{3\pi}{2}$}
\label{fig:sym_prob_amp}
\end{figure}

If we are to have a symmetric distribution of the probability amplitudes of the state, the middle two terms (corresponding to states $\ket{01}$ and $\ket{10}$ should be equal while the first and the last term (corresponding to $\ket{00}$ and $\ket{11}$) should be equal. That would give us the following.
\begin{equation}
\begin{aligned}
\cos{(\theta _1/2)}\sin{(\theta _2/2)} - \sin{(\theta _1/2)}\cos{(\theta _2/2)} = 0 \Rightarrow \sin {((\theta _2 - \theta _1)/2)} = 0 \\
\cos{(\theta _1/2)}\cos{(\theta _2/2)} - \sin{(\theta _1/2)}\sin{(\theta _2/2)} = 0 \Rightarrow \cos {((\theta _2 + \theta _1)/2)} = 0
\end{aligned}    
\end{equation}
The above condition can be satisfied under the following conditions.
\begin{flalign}
\begin{aligned}
\theta _2 &= 2n_1\pi + \theta _1 ~~ \textrm{for }n_1 = 0, 1, \hdots \\
\theta _2 &= (2n_2 + 1)\pi - \theta _1 ~~ \textrm{for }n_2 = 0, 1, \hdots 
\end{aligned}\label{sym_rel1}
\end{flalign}
From (\ref{sym_rel1}), we get the following expressions for $\theta _1$ and $\theta _2$.
\begin{flalign}
\begin{aligned}
\theta _1 &= \pm \frac{(2n + 1)\pi}{2} ~~ \textrm{for }n = 0, 1, \hdots \\
\theta _2 &= (2n_1\pi \pm \frac{(2n + 1)\pi}{2}) ~~ \textrm{for }n_1 = 0, 1, \hdots 
\end{aligned}\label{sym_rel2}
\end{flalign}
The above conditions will result in a distribution of probability amplitudes as given in Fig. \ref{fig:sym_prob_amp}.
In the above cases, although the distribution is symmetric, the probability distribution (amplitude squared) will be uniform with all states having identical probability values. In order to have distributions other than uniform, we could keep the value of $\theta _1$ as derived in (\ref{sym_rel2}) and play on the values of $\theta _2$. This would essentially keep the $\cos{(\theta _1/2)}$ and $\sin{(\theta _1/2)}$ terms of the MSB equal at $\frac{1}{\sqrt{2}}$. The terms corresponding to the lower significant bits would then play a role in generating the variations in the distributions.

\subsection{Non-uniform symmetric distributions}
\begin{figure}[h]
\centering 
\includegraphics[width=4in]{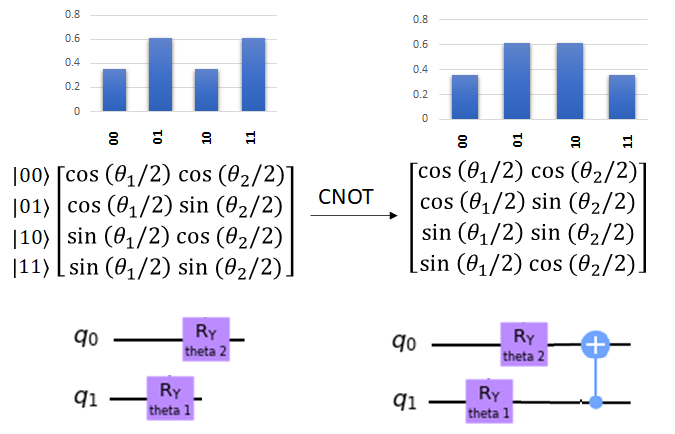} 
\caption{Flipping probability amplitudes of states for a symmetric distribution}
\label{fig:hy_2qb-2}
\end{figure}

In Fig. \ref{fig:hy_2qb-1}, we can see that if $\theta _1 =  \frac{\pi}{2}$ radians ($n=0$ in (\ref{sym_rel2})), the probability amplitudes corresponding to $\ket{00}$ and $\ket{10}$ will become equal with a value $\frac{1}{\sqrt{2}}\cos{(\theta _2/2)}$, while those corresponding to $\ket{01}$ and $\ket{11}$ will also become equal with a value $\frac{1}{\sqrt{2}}\sin{(\theta _2/2)}$. To have a probability distribution symmetric around the centre, we could exchange the probability amplitudes of $\ket{10}$ and $\ket{11}$, as shown in Fig. \ref{fig:hy_2qb-2}. 
The exchange of the probability amplitudes of the states $\ket{10}$ and $\ket{11}$ can be achieved by the application of an $X$ Pauli gate, conditioned on the first qubit (MSB) being in state $\ket{1}$. This is the CNOT gate as shown in Fig. \ref{fig:hy_2qb-2}. The bar plot in the figure is for a $\theta _2 = \frac{2\pi}{3}$ radians. A CNOT conditioned on $\ket{0}$ in the first qubit would have resulted in a symmetric distribution where the corner amplitudes are higher than the amplitudes at the centre (exchange of probability amplitudes of the states $\ket{00}$ and $\ket{01}$). To generate distributions that have central tendencies, we will first have to figure a suitable value for the rotation angles of the lower significant bits, and then have to decide the control state of the MSB over which the states are flipped. To generate the kind of distribution seen in Fig. \ref{fig:hy_2qb-2}, we must have $\cos{(\theta _2/2)} < \sin{(\theta _2/2)}$. In other words, we must have $\frac{5\pi}{2} > \theta _2 > \frac{\pi}{2}$.

\begin{figure}[h]
\centering 
\includegraphics[width=5in]{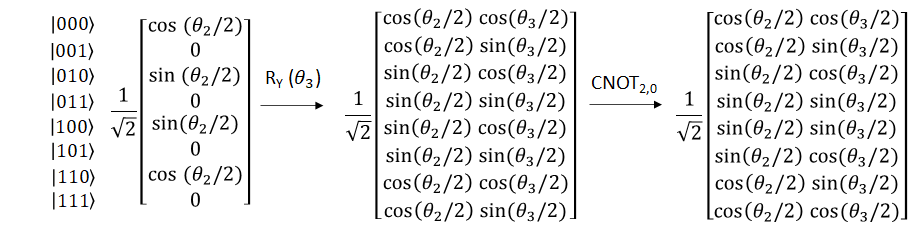} 
\caption{State amplitudes after the addition of a qubit}
\label{fig:hy_2qb-4}
\end{figure}

\begin{figure}[h]
\centering 
\includegraphics[width=5in]{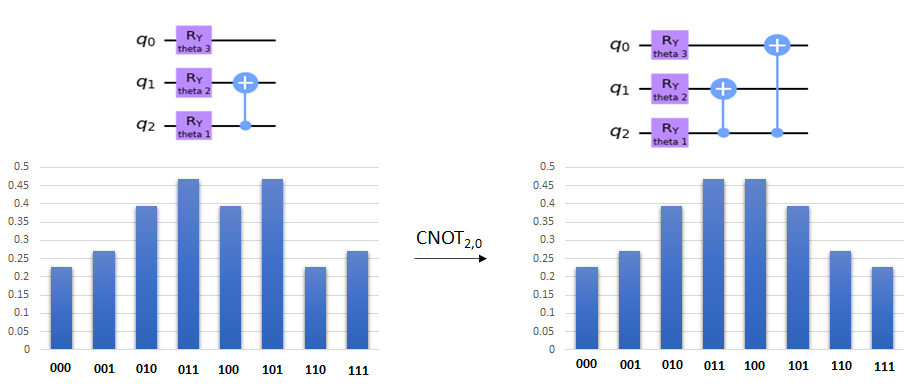} 
\caption{Distribution of probability amplitudes after the application of the CNOT gate}
\label{fig:hy_2qb-5}
\end{figure}

Let us now add a third qubit to the system. The numbering scheme of the qubits will then change. The qubit corresponding to the MSB will become $q_2$. The lower significant bits will then become $q_1$ and $q_0$. The third qubit is then passed through a rotation gate $R_Y$ with parameter $\theta _3$. Addition of the third qubit will result in the expansion of the number of states to eight. The $R_Y (\theta _3)$ operation on the third qubit will result in each of the four existing states of the 2 qubit system dividing up with an extra $\cos{(\theta _3/2)}$ and $\sin{(\theta _3/2)}$ terms as shown in Fig. \ref{fig:hy_2qb-4}.
In Fig. \ref{fig:hy_2qb-4}, the $\cos{(\theta _1/2)}$ and $\sin{(\theta _1/2)}$ terms have been taken out as $\frac{1}{\sqrt(2)}$, since $\theta _1 = \frac{\pi}{2}$. If we look at the states after the application of $R_Y (\theta _3)$ to the third qubit, we can see that if the amplitudes of the states $\ket{100}$ and $\ket{101}$ followed by $\ket{110}$ and $\ket{111}$ are exchanged, the distribution of the amplitudes are symmetrical. This exchange can happen if conditioned on the qubit corresponding to the MSB, i.e., $q_2$, being in state $\ket{1}$, a CNOT gate is applied to the qubit corresponding the lowest significant bit (LSB), i.e., $q_0$. This is depicted in Fig. \ref{fig:hy_2qb-5}.
The bar plot in Fig. \ref{fig:hy_2qb-5} was created with $\theta _2 = \frac{2\pi}{3} (120^{\circ})$ and $\theta _3 = 100^{\circ}$.  From the figure it is clear that by exchanging the amplitudes of states $\ket{100}$ and $\ket{101}$ and further in $\ket{110}$ and $\ket{111}$, a symmetric distribution can be achieved.

The distribution as seen in Fig. \ref{fig:hy_2qb-5} resembles a Gaussian distribution. This is a distribution that has monotonically decreasing amplitudes on either side of the centre. Not all symmetric distributions need to have this feature. However, if we are to model such distributions, it is important that we find the criteria that the parameters ($\theta _2$ and $\theta _3$ in this case) must satisfy. Let us look at the amplitudes where MSB is $0$. If we look at the states $\ket{000}$ and $\ket{001}$, the amplitude of $\ket{001}$ will be more than $\ket{000}$, only if $\cos{(\theta _3/2)}\leq \sin{(\theta _3/2)}$. This condition will also ensure that the amplitude of $\ket{011}$ is more than the amplitude of $\ket{010}$. Another condition that needs to be satisfied is that the amplitude of $\ket{010}$ is more than the amplitude of $\ket{001}$. That will happen if the following condition is satisfied.
\begin{flalign*}
\sin{(\theta _2/2)}\cos{(\theta _3/2)} & \geq \cos{(\theta _2/2)}\sin{(\theta _3/2)} \\
\Rightarrow ~~ \sin{((\theta _2 - \theta _3)/2)} &\geq 0 
\end{flalign*}
To put it all together, the conditions that the parameters need to obey such that we have a distribution which has a central tendency with a monotonically decreasing function on either side, are given as follows.
\begin{flalign*}
\begin{aligned}
\frac{\pi}{2} \leq \theta _3 \leq \frac{5\pi}{2} \\
0 \leq \theta _2 - \theta _3 \leq 2\pi
\end{aligned} 
\end{flalign*}
For an $n$ qubit system, we will need to satisfy the constraints given in (\ref{eqn:param_cond}).
\begin{flalign}
\begin{aligned}
\frac{\pi}{2} \leq & \theta _k \leq \frac{5\pi}{2} ~~~ k = 2, 3, \hdots n\\
0 \leq \theta _j &- \theta _{j+1} \leq 2\pi ~~~ j = 2, 3, \hdots n-1\\
\end{aligned} \label{eqn:param_cond}
\end{flalign}

To summarize, to have a symmetric distribution we need to have $\theta _1 = \frac{\pi}{2}$ and circuits of the type given in Fig. \ref{fig:hy_2qb-5}. To generate distributions that have a central tendency, like in Gaussian distribution curves, constraints of the type given in (\ref{eqn:param_cond}) need to be met. If we are to match the generated distribution curve to a given distribution curve, we could use variational solvers to do that. This is discussed in section \ref{Vsolver}. For an $n$ qubit system, the above circuit construction requires $n$ $R_Y$ rotation gates and $n-1$ CNOT gates. When used with a variational solver, we will need $n-1$ parameters. 

\subsection{Asymmetric distribution}
When we say asymmetric distributions, we mean distributions that have skewness in the distributions. It could be distributions like skew-normal, log-normal distributions etc. We approach this problem by dividing the distribution into two areas, based on the MSB. Positive skewness indicates that the mode of the distribution lies on the side where the qubit corresponding to the MSB has the state $\ket{0}$. The opposite ($\ket{1}$) is true for negative skewness. Distributions with positive skewness will have the monotonically decreasing tail on the side where the MSB qubit has state $\ket{1}$. This part (MSB with state $\ket{1}$) can be handled by following the approach given in (\ref{eqn:param_cond}). However, the inequalities in (\ref{eqn:param_cond}) will get reversed as given in (\ref{eqn:param_cond-asym1}). Conditioned on the MSB qubit in state $\ket{1}$, the parameter values of the $R_Y$ rotation gates on the lower significant bits are progressively increased. 
\begin{flalign}
\begin{aligned}
-\frac{3\pi}{2} \leq & \theta _k \leq \frac{\pi}{2} ~~~ k = 2, 3, \hdots n\\
0 \leq \theta _{j+1} &- \theta _{j} \leq 2\pi ~~~ j = 2, 3, \hdots n-1\\
\end{aligned} \label{eqn:param_cond-asym1}
\end{flalign}
It is to be noted here that, $\theta _1$ will depend upon the nature of the skewness. A positive skewness would require $\theta _1 \leq \frac{\pi}{2}$ ($\cos{\theta _1} \geq \sin{\theta _1}$). For negative skewness the inequalities need to be reversed. 

\begin{figure}[h]
\centering 
\includegraphics[width=5in]{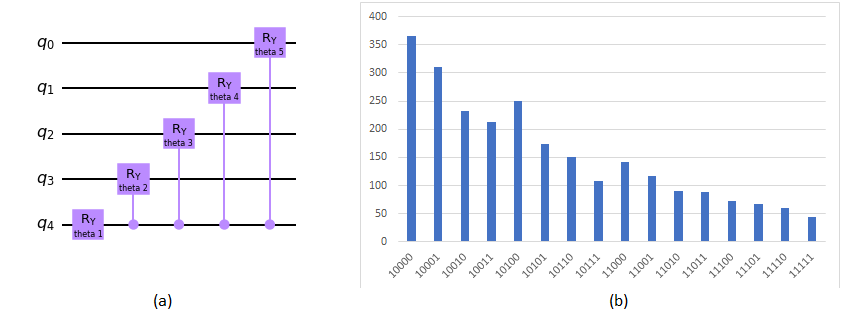} 
\caption{Generating the tail side of the asymmetric distribution where MSB qubit is in state $\ket{1}$}
\label{fig:hy_2qb-6}
\end{figure}

Fig. \ref{fig:hy_2qb-6}(a) gives the circuit of such an implementation in a 5 qubit system. Fig. \ref{fig:hy_2qb-6}(b) gives the count statistics of the states that have MSB qubit as $\ket{1}$, after the execution of the circuit for 10000 runs in a Qasm simulator \cite{qiskit}. The angles considered for the circuit were $\theta _1 = \pi/3 ~(60^{\circ}), ~\theta _2 = 0.361\pi ~(65^{\circ}), ~\theta _3 = 0.416\pi ~(75^{\circ}), ~\theta _4 = 0.444\pi ~(80^{\circ}), ~\theta _5 = 0.472\pi ~(85^{\circ})$.

\begin{figure}[h]
\centering 
\includegraphics[width=3in]{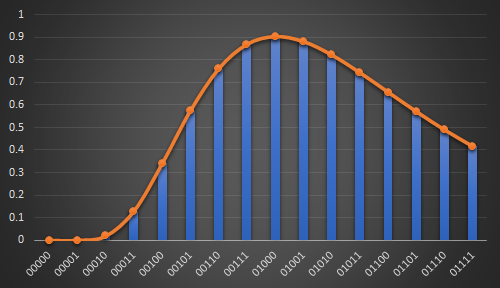} 
\caption{The side of a log-normal distribution where the MSB qubit is in state $\ket{0}$}
\label{fig:logn_dist}
\end{figure}

\begin{figure}[h]
\centering 
\includegraphics[width=6in]{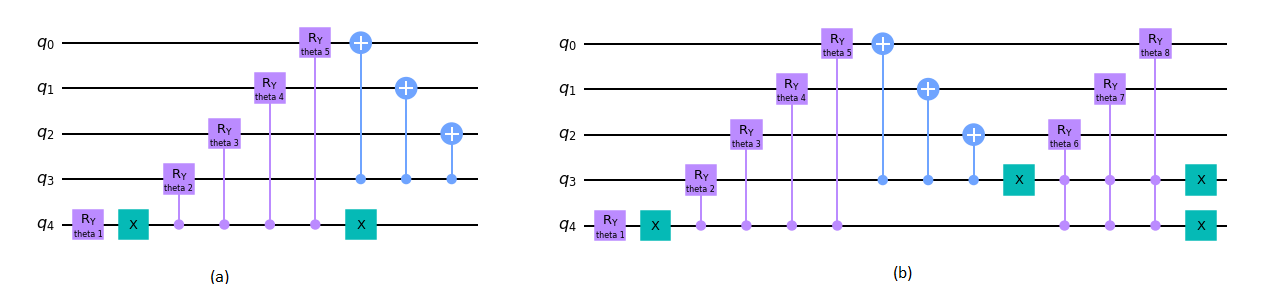} 
\caption{Circuit construction for the side where MSB qubit is in state $\ket{0}$}
\label{fig:hy_2qb-8}
\end{figure}

The other side of the distribution pertaining to the MSB qubit at state $\ket{0}$, will have the state with the peak probability amplitude. It will also have adjacent states with falling amplitudes. Fig.\ref{fig:logn_dist} shows the part of a log-normal distribution ($\mu = 0, \sigma = 0.5$) where the MSB qubit has state $\ket{0}$. The distribution is for a 5 qubit system having a total of 32 data points (16 with MSB as $\ket{0}$ and 16 with MSB as $\ket{1}$). 
From the figure, it can be seen that there is some amount of symmetry here. In the symmetric distribution, we had seen the symmetry is around the median and peak amplitude. This is also the place where the the MSB qubit state changes from $\ket{0}$ to $\ket{1}$. In the asymmetric or the skew-symmetric case, however, the peak probability amplitude (around which there is some symmetry) can be found around the point where the second most significant qubit changes from state $\ket{0}$ to $\ket{1}$. We could then follow the method we used while generating symmetric distributions. This however has to be conditioned on the MSB qubit in state $\ket{0}$. Fig. \ref{fig:hy_2qb-8}(a) gives the circuit construction for a 5 qubit system. This is identical to the construction given in Fig. \ref{fig:hy_2qb-5}, except that all the $R_Y$ gates are now controlled and acts only when the MSB qubit is in state $\ket{0}$. The two $X$ gates ensure that the control is on $\ket{0}$.
Now that the symmetry is being considered only on one side, the control qubit for the CNOT gates should be the second MSB qubit, i.e., $q_3$ in Fig. \ref{fig:hy_2qb-8}(a). The constraints on the angles could be as given in (\ref{eqn:param_cond}), except that the indices $j$ and $k$ will now start from 3 instead of 2. Note that we are not explicitly making the assignment $\theta _2 = \frac{\pi}{2}$, since the distribution is not an exact symmetry. The finer adjustments and estimation of the rotation angle parameters may be done by the variational solver. 

\begin{figure}[h]
\centering 
\includegraphics[width=6in]{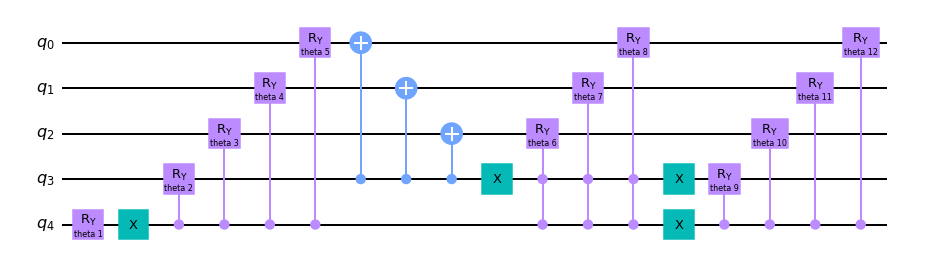} 
\caption{Entire circuit construction in a 5 qubit system for an asymmetric distribution}
\label{fig:hy_2qb-9}
\end{figure}

The construction in Fig. \ref{fig:hy_2qb-8}(a) would mainly lead to a somewhat symmetric kind of a distribution. To bring in an extra layer of asymmetry we could add some more $R_Y$ gates but controlled on the two most significant qubits, i.e. $q_4$ and $q_3$ in a 5 qubit system, as shown in Fig. \ref{fig:hy_2qb-8}(b). In the figure, the two $X$ gates on $q_3$ ensure that the extra layer of controlled rotations happen only when the two most significant qubits are in the state $\ket{00}$. This will fine tune the probability amplitudes for the states $\ket{00000}, \ket{00001} \hdots \ket{00111}$. We could very well do these kind of fine tuning when the two most significant qubits are in the state $\ket{01}$. This will be dependent on the type of distribution that we are trying to match or train our circuit for. A highly skewed distribution may require either/or of such fine adjustments. Our final circuit construction for the asymmetric distribution is given in Fig. \ref{fig:hy_2qb-9}.
For a $n$ qubit system, this circuit construction would need the following. 
\begin{itemize}
    \item 1 $R_Y$ rotation gate,
    \item $2(n-1)$ controlled $R_Y$ rotation gates,
    \item $n-2$ doubly controlled rotation gates,
    \item $n-2$ CNOT gates,
    \item $4$ Pauli $X$ gates.
\end{itemize}
A variational solver for this circuit would need $3n-3$ parameters.

\subsubsection{Tackling stronger skewness}\label{skew-dist}
\begin{figure}[h]
\centering 
\includegraphics[width=6in]{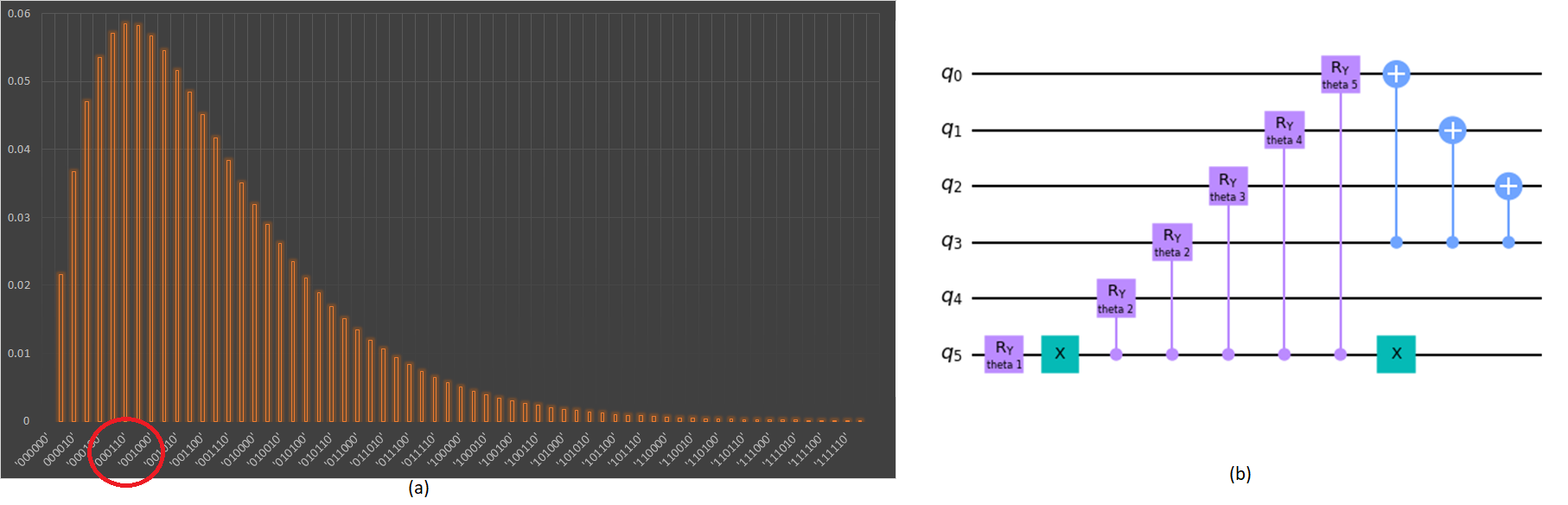} 
\caption{(a) Chi-square ($k=4$) distribution discretized over six qubit states (b) Circuit construction for generating the distribution where MSB qubit is in state $\ket{0}$}
\label{fig:hy_2qb-11}
\end{figure}

In the example where we discussed a log-normal discussion, the peak amplitude is around the point where the second most significant qubit changes states. We could have distributions where the skewness is even stronger and the peak amplitude can be found where a lower significant qubits changes state, as shown in Fig. \ref{fig:hy_2qb-11}(a).
In Fig. \ref{fig:hy_2qb-11}(a), a chi-square distribution with degrees of freedom, $k=4$, is shown. The distribution is discretized over 64 ($2^6$) points and shown. From the figure it can be seen that the peak amplitude occurs where the third most significant qubit changes state ($\ket{000111}$ to $\ket{001000}$). We will need symmetry around this point. This can be achieved by having CNOT gates at the lower significant qubits controlled at the third qubit. This is shown in Fig. \ref{fig:hy_2qb-11}(b). The circuit construction is similar to what we have in Fig. \ref{fig:hy_2qb-8}(a), except for the CNOT gates where the control has shifted. For any given distribution the control will shift based on the skewness and the point where the peak amplitude is located.
Like we had for the symmetric distribution in (\ref{eqn:param_cond}), we will need the following constraints here. 
\begin{flalign}
\begin{aligned}
\frac{\pi}{2} \leq & \theta _k \leq \frac{5\pi}{2} ~~~ k = 4, 5, \hdots n\\
0 \leq \theta _j &- \theta _{j+1} \leq 2\pi ~~~ j = 4, 5, \hdots n-1\\
\end{aligned} \label{eqn:param_cond_skew}
\end{flalign}

The circuit construction will remain the same as given in Fig. \ref{fig:hy_2qb-6} for generating the tail side of the distribution where the MSB qubit is in state $\ket{1}$.

\subsection{Drawbacks to the tracing methodology}
While the technique proposed here may be able to generate distributions that are close to the actual desired ones, there are some drawbacks to it. One primary drawback to this method is the presence of CNOT and $R_Y$ gates controlled on some significant qubit. In a large circuit system having a number of qubits, implementing this kind of an architecture in a real hardware will require a lot of swap gates. The most significant qubits will have to be positioned in the hardware in such a way that it has the maximum number of connections with other qubits. Wherever, connections are lacking, swap gates come into the picture resulting in increased gate counts and circuit depth. In other words, we will need a highly interconnected hardware to scale the method proposed here. The circuit proposed in \cite{Zoufal} does not suffer from this drawback as the controlled gates require adjacent connections. 

The second major drawback of this method is the number of controlled gates used. Controlled gates like CNOT, controlled $R_Y$ gates are prone to noise. CNOT gates also come into the picture when swap gates are used. With a large number of such gates, the results obtained will have errors. This is especially true for the circuits generating asymmetric distributions. If we are to implement a larger system (more states representing probability distributions), in a real hardware, we will need gate noise error mitigation methods to be implemented as well. 

\section{Using a variational solver to generate desired distributions} \label{Vsolver}
\begin{figure}[h]
\centering 
\includegraphics[width=4in]{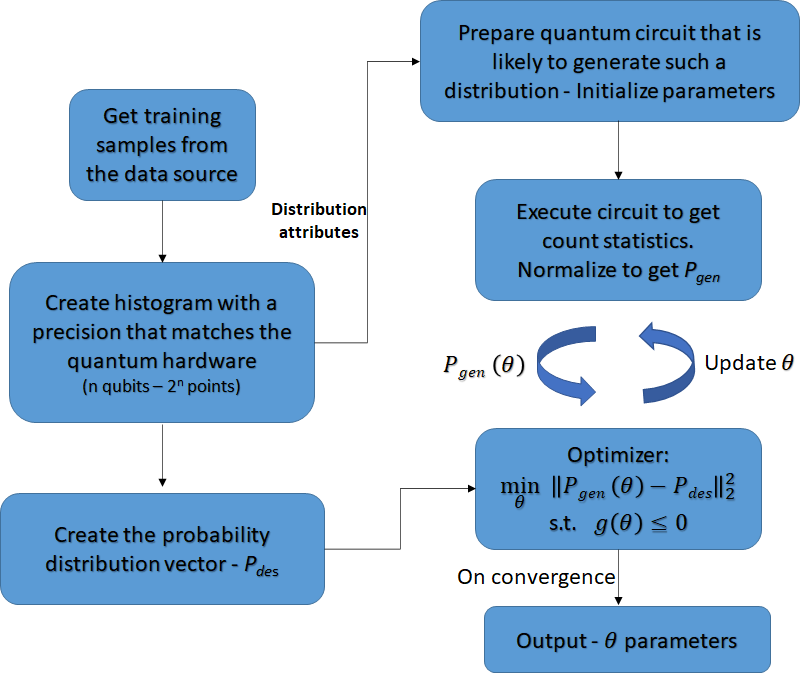} 
\caption{Variational solver to optimize the circuit parameters of a distribution generator}
\label{fig:var_sol}
\end{figure}

The objective of this exercise could be to train a quantum circuit with some given data sampled from an unknown distribution. If some properties of the distribution is known (location of the mean/median/mode, symmetric/asymmetric, skewness, etc.), the circuits can be suitably modified and the parameters can be optimized by a variational solver \cite{Peruzzo}, \cite{Zoufal}. The optimized parameterized quantum circuit (PQC) will then be able to generate samples that could mimic the given distribution. Fig. \ref{fig:var_sol} gives the schematic of the variational solver. 

The variational solver will have two parts. A quantum computation part and a classical computation part. The quantum part of the solver is where the execution of the circuit happens. The count statistics of the states are collated and normalized to get the probability distribution. The classical computation part involves the optimizer where the parameters $\theta$ are updated subject to the constraints given in $g(\theta)$. The updated parameters are sent back to the quantum circuit, where the circuit is updated and executed again. This cycle is continued till convergence. The final set of parameters can be used to generate samples. 

\section{Results and discussion}
\begin{figure}[!h]
\centering 
\includegraphics[width=5.8in]{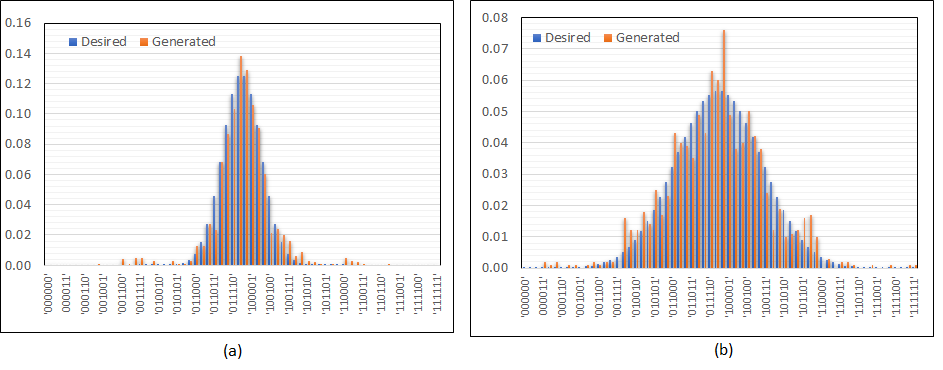} 
\caption{Generated Gaussian distributions in a qasm simulator, 6 qubit system, normalized (a) $e^{-0.5x^2}$, (b) $e^{-0.1x^2}$}
\label{fig:gen_normal}
\end{figure}

Before we look at some metric to check how the circuits perform to generate known symmetric and asymmetric distributions, let us first visually compare the generated distributions with the real ones. We first look at Gaussian distributions with very different variances. Fig. \ref{fig:gen_normal} gives the bar plot of actual (desired) versus generated distributions in a qasm simulator using a circuit consisting of 6 qubits.
The x-axis in the plots are the states of the 6 qubit system. $x$ ranges from $\left[ -10, 10 \right]$. We see a good match between the generated and the desired distributions. We used the COBYLA optimizer in the variational solver \cite{qiskit}. The constraints given in (\ref{eqn:param_cond}) were used in the optimization process. The variational solver can generate distributions over all types of variances as long as the constraints are satisfied.

We now look at an asymmetric distribution. We have used a log-normal distribution for comparison. We used the two different circuits shown in Fig. \ref{fig:hy_2qb-8}. In Fig. \ref{fig:hy_2qb-8}(a), the circuit is a controlled version (controlled on MSB qubit) of the circuit shown in Fig. \ref{fig:hy_2qb-5}, when the most significant qubit has state $\ket{0}$. In Fig. \ref{fig:hy_2qb-8}(b), the circuit has extra rotation gates to fine tune the distribution. In this case, the constraints given in (\ref{eqn:param_cond-asym1}) were used in the optimization process.
\begin{figure}[h]
\centering 
\includegraphics[width=5.8in]{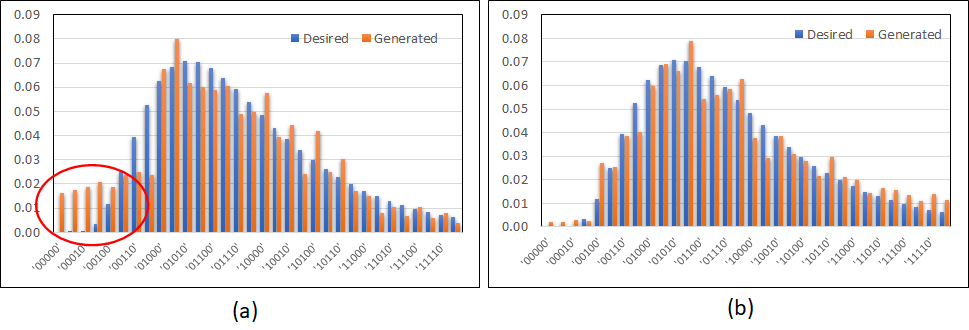} 
\caption{Generated log-normal distribution in a qasm simulator, 5 qubit system, normalized $e^{-2\log{x}^2}$  (a) controlled symmetry, (b) with added controlled rotation gates. $x \in [0, 3]$}
\label{fig:gen_lognormal}
\end{figure}

Fig. \ref{fig:gen_lognormal} gives the bar plot of the log-normal distributions using the two different circuits discussed above. In Fig. \ref{fig:gen_lognormal}(a), one can see that there is a discrepancy between the generated and the desired distribution at the point where the two most significant qubits have states $\ket{0}$. This is shown in the figure by a red circle. This discrepancy, however, goes away on the addition of the added rotation gates.

\begin{figure}[h]
\centering 
\includegraphics[width=5in]{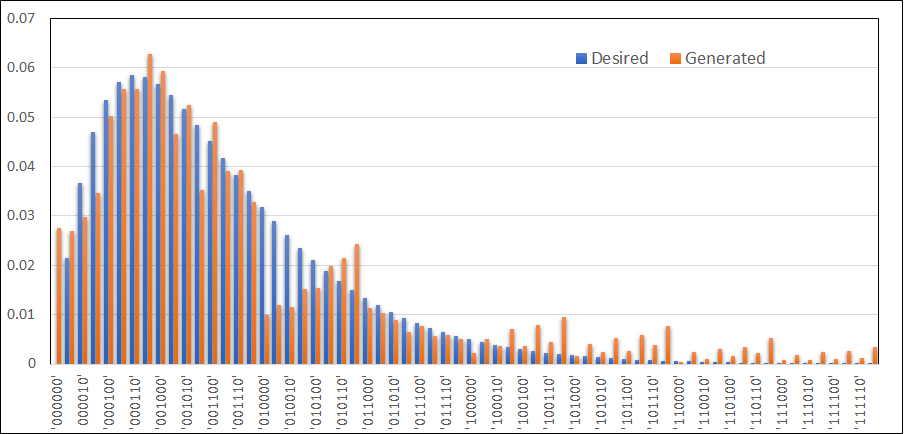} 
\caption{Generated chi-square distribution ($k=4$) in a qasm simulator, 6 qubit system,  $x \in [0, 20]$}
\label{fig:asym_chi}
\end{figure}

We also simulated for generating highly skewed distributions. Fig. \ref{fig:asym_chi} gives the bar plots of actual versus generated chi-square distribution. The distribution was generated using the circuit construction elaborated in section \ref{skew-dist}. The simulation was done in a qasm simulator.

We look at some metrics to see how the quantum circuits perform with respect to the variational circuit proposed in \cite{Zoufal} and as shown in Fig. \ref{fig:Zoufal_var}. In \cite{Zoufal}, the objective was to train a parameterized quantum circuit to generate distributions that resemble the distribution of a training data-set. The proposed method uses a QGAN, as elaborated in the related art section. For the sake of comparison we only use the variational circuit proposed in the paper and not the entire algorithm involving a QGAN. We check the output distribution of this circuit when several layers of the rotation gates and entanglers are repeated. We refer to this variational circuit by the term Ry-CZ VC. We refer to the circuits proposed here as Adaptive VC. The algorithm to generate distributions using the Adaptive VC is given in Fig. \ref{fig:var_sol}. The metrics used for comparisons are, 
\begin{enumerate}
    \item relative entropy,
    \item the second norm of the difference between the generated and the desired distributions, 
    \item Kolmogorov-Smirnov (KS) test statistics.
\end{enumerate}

Table \ref{table:sym_comparisons} gives the values of these metrics for both the Ry-CZ VC as well as the Adaptive VC for a desired distribution given by normalized $e^{-0.1x^2}$. While trying out with Ry-CZ VC, the effect of having multiple layers was also noted. 
\begin{table}[!h]
\caption{Results with symmetric distribution in a qasm simulator}
\vspace{-0.2cm}
\begin{center}
\begin{tabular}{|c|c|c|c|c|c|c|c|}
 \hline
\multirow{2}{*}{Qubits}  & \multirow{2}{*}{Layers} & \multicolumn{2}{|c|}{Rel. Entropy} & \multicolumn{2}{|c|}{$\lVert P_{gen} - P_{des} \Vert _2 ^2$} & \multicolumn{2}{|c|}{KS test - p value}\\ 
 \cline{3-8}
 & & Ry-CZ & Adaptive & Ry-CZ & Adaptive & Ry-CZ & Adaptive \\
 \hline
 \multirow{3}{*}{5} & 1	& 0.1765 & \multirow{3}{*}{0.0535} & 0.1087 & \multirow{3}{*}{0.0477} & 0.2355 & \multirow{3}{*}{0.3405}\\
  & 2 & 0.3061 & & 0.1115 & & 0.2402 & \\
  & 3 & 0.3034 & & 0.1059 & & 0.3551 & \\
  \hline
 \multirow{3}{*}{6} & 1	& 0.3577 & \multirow{3}{*}{0.0725} & 0.1042 & \multirow{3}{*}{0.0449} & 0.0586 & \multirow{3}{*}{0.0641}\\
  & 2 & 0.2952 & & 0.0930 & & 0.1105 & \\
  & 3 & 0.5689 & & 0.1177 & & 0.1079 & \\
  \hline
\end{tabular} \label{table:sym_comparisons}
\end{center}
\end{table} 
The results given in Table \ref{table:sym_comparisons} are the average over multiple runs. 

Table \ref{table:asym_comparisons} gives the values of the metrics for a desired distribution given by normalized $e^{-2\log{x}^2}$.
\begin{table}[!h]
\caption{Results with asymmetric (log-normal) distribution in a qasm simulator}
\vspace{-0.2cm}
\begin{center}
\begin{tabular}{|c|c|c|c|c|c|c|c|}
 \hline
\multirow{2}{*}{Qubits}  & \multirow{2}{*}{Layers} & \multicolumn{2}{|c|}{Rel. Entropy} & \multicolumn{2}{|c|}{$\lVert P_{gen} - P_{des} \Vert _2 ^2$} & \multicolumn{2}{|c|}{KS test - p value}\\ 
 \cline{3-8}
 & & Ry-CZ & Adaptive & Ry-CZ & Adaptive & Ry-CZ & Adaptive \\
 \hline
 \multirow{3}{*}{5} & 1	& 0.2120 & \multirow{3}{*}{0.0503} & 0.09 & \multirow{3}{*}{0.0369} & 0.7239 & \multirow{3}{*}{0.9141}\\
  & 2 & 0.1912 & & 0.0830 & & 0.6718 & \\
  & 3 & 0.3624 & & 0.0961 & & 0.6726 & \\
  \hline
 \multirow{3}{*}{6} & 1	& 0.2963 & \multirow{3}{*}{0.0953} & 0.0558 & \multirow{3}{*}{0.0366} & 0.3617 & \multirow{3}{*}{0.6813}\\
  & 2 & 0.3584 & & 0.0749 & & 0.5041 & \\
  & 3 & 0.7365 & & 0.0954 & & 0.6033 & \\
  \hline
\end{tabular} \label{table:asym_comparisons}
\end{center}
\end{table} 
It is clear from the KS statistics in the tables that the similarities between the desired and the generated distributions are statistically significant. This is true for both the Ry-CZ VC as well as the Adaptive VC. In the other metrics, the Adaptive VC appears to be doing better than the Ry-CZ VC. The low values of the relative entropy suggests that the generated distributions are a close match to the desired ones. In the Ry-CZ VC, circuits with a maximum of 2 layers is sufficient, and a third layer may not add more value to the distributions.

A similar comparison was also done for the highly skewed chi-square distribution discussed above. Table \ref{table:skew_comparisons} gives the results.
\begin{table}[!h]
\caption{Results with chi-square distribution in a qasm simulator}
\vspace{-0.2cm}
\begin{center}
\begin{tabular}{|c|c|c|c|c|c|c|c|}
 \hline
\multirow{2}{*}{Qubits}  & \multirow{2}{*}{Layers} & \multicolumn{2}{|c|}{Rel. Entropy} & \multicolumn{2}{|c|}{$\lVert P_{gen} - P_{des} \Vert _2 ^2$} & \multicolumn{2}{|c|}{KS test - p value}\\ 
 \cline{3-8}
 & & Ry-CZ & Adaptive & Ry-CZ & Adaptive & Ry-CZ & Adaptive \\
 \hline
 \multirow{3}{*}{6} & 1	& 0.4344 & \multirow{3}{*}{0.1496} & 0.0749 & \multirow{3}{*}{0.0567} & 0.5113
 & \multirow{3}{*}{0.1507}\\
  & 2 & 0.3319 & & 0.0753 & & 0.3647 & \\
  & 3 & 0.4228 & & 0.0856 & & 0.4483 & \\
  \hline
\end{tabular} \label{table:skew_comparisons}
\end{center}
\end{table} 
Here again, we see the Adaptive VC performing better. The results, however, are not as good as the previous two cases.

\begin{figure}[h]
\centering 
\includegraphics[width=6in]{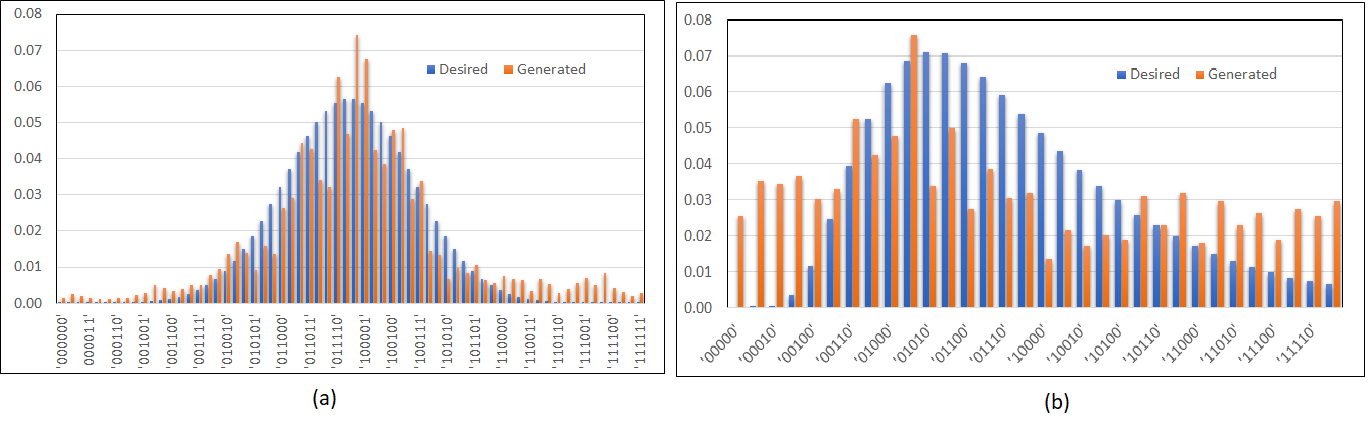} 
\caption{Execution in ibm\_hanoi. Normalized (a) $e^{-0.1x^2}$ in a 6 qubit system, (b) $e^{-2\log{x}^2}$ in a 5 qubit system.}
\label{fig:sym_hanoi}
\end{figure}

We ran the circuits in a real IBM hardware in the platform provided by IBM Quantum services \cite{ibm-quantum}. In this paper we used ibm\_hanoi and ibm\_auckland, which are one of the IBM Quantum Falcon processors. We used the Qiskit Runtime sampler primitive to execute the circuit and generate the distributions \cite{runtime}. Table \ref{table:hardware} gives the results. We used 2 layers in the Ry-CZ VC, when executing with IBM machines.
\begin{table}[!h]
\caption{Results with IBM Quantum machines}
\vspace{-0.2cm}
\begin{center}
\begin{tabular}{|c|c|c|c|c|c|}
 \hline
\multirow{2}{*}{Dist.} & \multirow{2}{*}{Qubits}  & \multicolumn{2}{|c|}{Rel. Entropy} & \multicolumn{2}{|c|}{$\lVert P_{gen} - P_{des} \Vert _2 ^2$} \\ 
 \cline{3-6}
 & & Ry-CZ & Adaptive & Ry-CZ & Adaptive \\
 \hline
 Gaussian & 5 & 0.3089 & 0.1243 & 0.1412 & 0.0810 \\
 Gaussian & 6 & 0.1834 & 0.1262 & 0.0635 & 0.0547 \\
 Log-normal & 5 & 0.1380 & 0.2424 & 0.0719 & 0.1013 \\
  \hline
\end{tabular} \label{table:hardware}
\end{center}
\end{table} 
Fig. \ref{fig:sym_hanoi} gives the bar plot of desired versus generated of a Gaussian distribution and a log-normal distribution, sampled in ibm\_hanoi. The Gaussian distribution generated by the quantum hardware is quite close to the expected distribution. However, we do see some deficiencies in the log-normal distribution as generated by the quantum hardware.

\section{Conclusions}
In this paper, we have given methodologies to construct parameterized quantum circuits that generate probability distributions. The proposed methodologies could be used as a general framework to build different kinds of distributions. We discussed both symmetric as well as asymmetric distributions. We used a variational scheme to optimize the parameters that reduce the distance between the generated and the desired distributions. We also suggest ways to fine tune the circuits such that the final distribution is close to the desired ones. We provided results to show that the circuits perform well to generate the distributions. Both simulators as well as hardwares were used to generate the distributions. We also compared our results with the ones generated by the Ry-CZ circuits proposed in \cite{Zoufal}. 

Although, the circuits proposed here performs well when compared with the Ry-CZ circuits, there are some limitations when the system is scaled to higher number of qubits. The circuits proposed here requires some amount of connectivity between the qubits, which may not be readily available in the hardwares. This would result in usage of a large number of swap gates and eventual increase in the number of CNOT gates. We also have a number of controlled gates in the proposed method. All these factors may impact the final result as noise creeps in. We will need error mitigation methods in our sampler to yield accurate results.

\bibliographystyle{abbrv}
\bibliography{ref}
\end{document}